\documentclass[prd, twocolumn, floatfix, nofootinbib, superscriptaddress]{revtex4-1}

\usepackage{graphicx}
\usepackage{amsmath}
\usepackage{amssymb}
\usepackage{hyperref}

\usepackage{color}
\usepackage{siunitx}
\usepackage{lineno}
\usepackage{textcase}

\begin{document}

\title{Explanation of the seasonal variation of cosmic multiple muon events observed with the NOvA Near Detector}

\newcommand{\ANL}{Argonne National Laboratory, Argonne, Illinois 60439, 
USA}
\newcommand{\Bandirma}{Bandirma Onyedi Eyl\"ul University, Faculty of 
Engineering and Natural Sciences, Engineering Sciences Department, 
10200, Bandirma, Balıkesir, Turkey}
\newcommand{\ICS}{Institute of Computer Science, The Czech 
Academy of Sciences, 
182 07 Prague, Czech Republic}
\newcommand{\IOP}{Institute of Physics, The Czech 
Academy of Sciences, 
182 21 Prague, Czech Republic}
\newcommand{\Atlantico}{Universidad del Atlantico,
Carrera 30 No.\ 8-49, Puerto Colombia, Atlantico, Colombia}
\newcommand{\BHU}{Department of Physics, Institute of Science, Banaras 
Hindu University, Varanasi, 221 005, India}
\newcommand{\UCLA}{Physics and Astronomy Department, UCLA, Box 951547, Los 
Angeles, California 90095-1547, USA}
\newcommand{\Caltech}{California Institute of 
Technology, Pasadena, California 91125, USA}
\newcommand{\Cochin}{Department of Physics, Cochin University
of Science and Technology, Kochi 682 022, India}
\newcommand{\Charles}
{Charles University, Faculty of Mathematics and Physics,
 Institute of Particle and Nuclear Physics, Prague, Czech Republic}
\newcommand{\Cincinnati}{Department of Physics, University of Cincinnati, 
Cincinnati, Ohio 45221, USA}
\newcommand{\CSU}{Department of Physics, Colorado 
State University, Fort Collins, CO 80523-1875, USA}
\newcommand{\CTU}{Czech Technical University in Prague,
Brehova 7, 115 19 Prague 1, Czech Republic}
\newcommand{\Dallas}{Physics Department, University of Texas at Dallas,
800 W. Campbell Rd. Richardson, Texas 75083-0688, USA}
\newcommand{\DallasU}{University of Dallas, 1845 E 
Northgate Drive, Irving, Texas 75062 USA}
\newcommand{\Delhi}{Department of Physics and Astrophysics, University of 
Delhi, Delhi 110007, India}
\newcommand{\JINR}{Joint Institute for Nuclear Research,  
Dubna, Moscow region 141980, Russia}
\newcommand{\Erciyes}{
Department of Physics, Erciyes University, Kayseri 38030, Turkey}
\newcommand{\FNAL}{Fermi National Accelerator Laboratory, Batavia, 
Illinois 60510, USA}
\newcommand{\FSU}{Florida State University, Tallahassee, Florida 32306, USA}
\newcommand{\UFG}{Instituto de F\'{i}sica, Universidade Federal de 
Goi\'{a}s, Goi\^{a}nia, Goi\'{a}s, 74690-900, Brazil}
\newcommand{\Guwahati}{Department of Physics, IIT Guwahati, Guwahati, 781 
039, India}
\newcommand{\Harvard}{Department of Physics, Harvard University, 
Cambridge, Massachusetts 02138, USA}
\newcommand{\Houston}{Department of Physics, 
University of Houston, Houston, Texas 77204, USA}
\newcommand{\IHyderabad}{Department of Physics, IIT Hyderabad, Hyderabad, 
502 205, India}
\newcommand{\Hyderabad}{School of Physics, University of Hyderabad, 
Hyderabad, 500 046, India}
\newcommand{\IIT}{Illinois Institute of Technology,
Chicago IL 60616, USA}
\newcommand{\Indiana}{Indiana University, Bloomington, Indiana 47405, 
USA}
\newcommand{\INR}{Institute for Nuclear Research of Russia, Academy of 
Sciences 7a, 60th October Anniversary prospect, Moscow 117312, Russia}
\newcommand{\UIowa}{Department of Physics and Astronomy, University of Iowa, 
Iowa City, Iowa 52242, USA}
\newcommand{\ISU}{Department of Physics and Astronomy, Iowa State 
University, Ames, Iowa 50011, USA}
\newcommand{\Irvine}{Department of Physics and Astronomy, 
University of California at Irvine, Irvine, California 92697, USA}
\newcommand{\Jammu}{Department of Physics and Electronics, University of 
Jammu, Jammu Tawi, 180 006, Jammu and Kashmir, India}
\newcommand{\Lebedev}{Nuclear Physics and Astrophysics Division, Lebedev 
Physical 
Institute, Leninsky Prospect 53, 119991 Moscow, Russia}
\newcommand{\Magdalena}{Universidad del Magdalena, Carrera 32 No 22-08 Santa Marta, Colombia}
\newcommand{\MSU}{Department of Physics and Astronomy, Michigan State 
University, East Lansing, Michigan 48824, USA}
\newcommand{\Crookston}{Math, Science and Technology Department, University 
of Minnesota Crookston, Crookston, Minnesota 56716, USA}
\newcommand{\Duluth}{Department of Physics and Astronomy, 
University of Minnesota Duluth, Duluth, Minnesota 55812, USA}
\newcommand{\Minnesota}{School of Physics and Astronomy, University of 
Minnesota Twin Cities, Minneapolis, Minnesota 55455, USA}
\newcommand{\Mississippi}{University of Mississippi, University, Mississippi 38677, USA}
\newcommand{\NISER}{National Institute of Science Education and Research, An OCC of Homi Bhabha
National Institute, Bhubaneswar, Odisha, India}
\newcommand{\OSU}{Department of Physics, Ohio State University, Columbus,
Ohio 43210, USA}
\newcommand{\Oxford}{Subdepartment of Particle Physics, 
University of Oxford, Oxford OX1 3RH, United Kingdom}
\newcommand{\Panjab}{Department of Physics, Panjab University, 
Chandigarh, 160 014, India}
\newcommand{\Pitt}{Department of Physics, 
University of Pittsburgh, Pittsburgh, Pennsylvania 15260, USA}
\newcommand{\QMU}{Particle Physics Research Centre, 
Department of Physics and Astronomy,
Queen Mary University of London,
London E1 4NS, United Kingdom}
\newcommand{\RAL}{Rutherford Appleton Laboratory, Science 
and 
Technology Facilities Council, Didcot, OX11 0QX, United Kingdom}
\newcommand{\SAlabama}{Department of Physics, University of 
South Alabama, Mobile, Alabama 36688, USA} 
\newcommand{\Carolina}{Department of Physics and Astronomy, University of 
South Carolina, Columbia, South Carolina 29208, USA}
\newcommand{\SDakota}{South Dakota School of Mines and Technology, Rapid 
City, South Dakota 57701, USA}
\newcommand{\SMU}{Department of Physics, Southern Methodist University, 
Dallas, Texas 75275, USA}
\newcommand{\Stanford}{Department of Physics, Stanford University, 
Stanford, California 94305, USA}
\newcommand{\Sussex}{Department of Physics and Astronomy, University of 
Sussex, Falmer, Brighton BN1 9QH, United Kingdom}
\newcommand{\Syracuse}{Department of Physics, Syracuse University,
Syracuse NY 13210, USA}
\newcommand{\Tennessee}{Department of Physics and Astronomy, 
University of Tennessee, Knoxville, Tennessee 37996, USA}
\newcommand{\Texas}{Department of Physics, University of Texas at Austin, 
Austin, Texas 78712, USA}
\newcommand{\Tufts}{Department of Physics and Astronomy, Tufts University, Medford, 
Massachusetts 02155, USA}
\newcommand{\UCL}{Physics and Astronomy Department, University College 
London, 
Gower Street, London WC1E 6BT, United Kingdom}
\newcommand{\Virginia}{Department of Physics, University of Virginia, 
Charlottesville, Virginia 22904, USA}
\newcommand{\WSU}{Department of Mathematics, Statistics, and Physics,
 Wichita State University, 
Wichita, Kansas 67260, USA}
\newcommand{\WandM}{Department of Physics, William \& Mary, 
Williamsburg, Virginia 23187, USA}
\newcommand{\Wisconsin}{Department of Physics, University of 
Wisconsin-Madison, Madison, Wisconsin 53706, USA}
\newcommand{\deceased}{Deceased.}
\affiliation{\ANL}
\affiliation{\Atlantico}
\affiliation{\Bandirma}
\affiliation{\BHU}
\affiliation{\Caltech}
\affiliation{\Charles}
\affiliation{\Cincinnati}
\affiliation{\Cochin}
\affiliation{\CSU}
\affiliation{\CTU}
\affiliation{\Delhi}
\affiliation{\Erciyes}
\affiliation{\FNAL}
\affiliation{\FSU}
\affiliation{\UFG}
\affiliation{\Guwahati}
\affiliation{\Houston}
\affiliation{\Hyderabad}
\affiliation{\IHyderabad}
\affiliation{\IIT}
\affiliation{\Indiana}
\affiliation{\ICS}
\affiliation{\INR}
\affiliation{\IOP}
\affiliation{\UIowa}
\affiliation{\ISU}
\affiliation{\Irvine}
\affiliation{\JINR}
\affiliation{\Magdalena}
\affiliation{\MSU}
\affiliation{\Duluth}
\affiliation{\Minnesota}
\affiliation{\Mississippi}
\affiliation{\OSU}
\affiliation{\NISER}
\affiliation{\Panjab}
\affiliation{\Pitt}
\affiliation{\QMU}
\affiliation{\SAlabama}
\affiliation{\Carolina}
\affiliation{\SMU}
\affiliation{\Sussex}
\affiliation{\Syracuse}
\affiliation{\Texas}
\affiliation{\Tufts}
\affiliation{\UCL}
\affiliation{\Virginia}
\affiliation{\WSU}
\affiliation{\WandM}
\affiliation{\Wisconsin}

\author{S.~Abubakar}
\affiliation{\Erciyes}

\author{M.~A.~Acero}
\affiliation{\Atlantico}

\author{B.~Acharya}
\affiliation{\Mississippi}

\author{P.~Adamson}
\affiliation{\FNAL}









\author{N.~Anfimov}
\affiliation{\JINR}


\author{A.~Antoshkin}
\affiliation{\JINR}


\author{E.~Arrieta-Diaz}
\affiliation{\Magdalena}

\author{L.~Asquith}
\affiliation{\Sussex}


\author{A.~Aurisano}
\affiliation{\Cincinnati}



\author{A.~Back}
\affiliation{\Indiana}
\affiliation{\ISU}



% ...

\collaboration{The NOvA Collaboration}
\noaffiliation

\date{\today}



\date{\today}
\begin{abstract}
    The flux of cosmic ray muons at the Earth's surface exhibits seasonal variations due to changes in the temperature of the atmosphere affecting the production and decay of mesons in the upper atmosphere. Using 1473 live days of data collected by the NuMI Off-axis $\nu_e$ Appearance (NOvA) Near Detector during 2018--2022, we studied the seasonal pattern in the multiple-muon event rate. The data confirm an anticorrelation between the multiple-muon event rate and effective atmospheric temperature, consistent across all the years of data. Previous analyses from MINOS and NOvA saw a similar anticorrelation but did not include an explanation. We find that this anticorrelation is driven by altitude--geometry effects as the average muon production height changes with the season. This has been studied with a CORSIKA cosmic ray simulation package by varying atmospheric parameters, and provides an explanation to a longstanding discrepancy between the seasonal phases of single and multiple-muon events.
\end{abstract}

\maketitle

\setlength\linenumbersep{5pt}
\section{Introduction}
The flux of muons produced in cosmic ray air showers and detected underground exhibits a well-known seasonal variation. This effect arises due to the competition between the decay and interaction of secondary mesons (\(\pi, K\)) produced in the primary cosmic-ray interaction. In the upper atmosphere, where most muons originate, higher temperatures in summer cause the atmosphere to expand, leading to lower density and a greater probability for mesons to decay into muons before interacting. Conversely, in winter, the atmosphere contracts, resulting in higher density, which increases the likelihood of mesons interacting before they can decay. As a result, the muon flux reaches a maximum in summer and a minimum in winter. In the past, this seasonal effect has been observed in multiple experiments~\cite{RevModPhys.24.133, sherman1954, seasonal1967, fenton1961, baksan1987, MACRO1997, amanda, Borexino, lvd2009, icecube, ssw2009, minosfd, minosnd, doublechooz2017, dayabay2018}.  For our case, since the NuMI Off-axis $\nu_e$ Appearance (NOvA) Near Detector (ND) is underground, the majority of the cosmic muons produced in the atmosphere do not reach the detector as the energy of the primary cosmic rays falls exponentially (\(\sim E^{-2.7}\)). Only a muon with an energy $E_\mu > \sec{\theta_{zen}} \times$ 54 GeV can reach the detector, where $\theta_{zen}$ is the muon zenith angle. 

Large neutrino detectors are capable of detecting multiple-muon events, where several nearly parallel muon tracks from a single cosmic ray air shower arrive at the detector. While the seasonal effect of single muons is well understood, the behavior of multiple-muon events presented an intriguing anomaly. The MINOS experiment first reported an unexpected winter maximum in multiple-muon events underground, contrasting with the well-established summer peak observed for single muons. This effect was observed in both the MINOS ND~\cite{minos}, located at 225 meters water equivalent (mwe), and the MINOS Far Detector (FD)~\cite{minos}, located at 2100 mwe underground. The MINOS study considered possible explanations, and favored the idea that secondary hadronic interactions in the atmosphere played an important role.  The NOvA ND~\cite{novaND} ($E_\mu~>$~54~GeV), located at the same depth as the MINOS ND but with a different detector design, confirmed the winter peak in multiple-muon events in a dataset spanning 2015–2017. To investigate this discrepancy, we conducted a detailed analysis of multiple-muon events using the NOvA ND. We concentrate on the size of the muon component of air showers compared to the size of our ND. 

In this paper, we extend the analysis of Ref.~\cite{novaND} using approximately four years of NOvA ND data (April 30, 2018 to May 11, 2022, corresponding to 1473 live days over four years). We explain the seasonal variation of multiple-muon events with the altitude-geometry effect, and check that explanation with a CORSIKA~\cite{corsika} based simulation of cosmic rays in the NOvA ND.

In Sec.~\ref{Sec: The NOvA Near Detector}, there are details of the NOvA ND. Section~\ref{sec:Cosmic muon simulation} describes the simulation and modeling by season of cosmic muons with CORSIKA. In Sec.~\ref{sec: Temperature Data} we summarize how we obtained temperature data. Next, Sec.~\ref{sec: Cosmic muon data in the NOvA Near Detector} presents the analysis of new multiple-muon data and an improved event rate calculation. Then, Sec.~\ref{sec:explanation} explains the observed seasonal behavior of the MINOS and NOvA multiple-muon rate as due to the altitude--geometry effect.  A method to determine whether this effect switches the phase of underground detectors as a function of detector size and depth is given.

\section{The NO\lowercase{v}A Near Detector}
\label{Sec: The NOvA Near Detector}

The NOvA experiment~\cite{novatdr} is a long-baseline neutrino oscillation experiment consisting of two functionally identical liquid-scintillator detectors. The ND, relevant for this analysis, is located at Fermilab approximately 105 m underground. In addition to neutrino interactions, the ND records a significant flux of cosmic-ray muons, making it well suited for studying seasonal variations of cosmic backgrounds. The ND has dimensions of \SI{4.0}{\meter} $\times$ \SI{4.0}{\meter} $\times$ \SI{15.9}{\meter}. It is divided into two regions: the active detector and the muon catcher. The active region consists of 192 planes, while the muon catcher contains 22 planes with each pair of scintillator planes followed by a \SI{10}{\centi\meter} thick steel absorber (except the last). 

The basic detector element is a polyvinyl chloride (PVC) cell of cross section \SI{3.9}{\centi\meter} $\times$ \SI{6.6}{\centi\meter} and length \SI{3.9}{\meter}, filled with liquid scintillator. A plane is formed by arranging these cells side by side. The planes are arranged alternately in
horizontal and vertical orientations to enable reconstruction of a 3D track.  

Scintillation light produced in each cell is collected by a wavelength-shifting fiber and transmitted to an avalanche photodiode (APD). The APD converts the light into an electronic signal, which is digitized by front-end electronics. Signals above a preset threshold are recorded as detector hits.


\section{Cosmic muon simulation}
\label{sec:Cosmic muon simulation}

Previous analyses of MINOS and NOvA did not compare their data with a cosmic simulation that included seasonal effects. For the first time, we used the CORSIKA simulation with seasonal effects incorporated in it. Cosmic muons are simulated with CORSIKA~\cite{corsika} v7.7410, using GHEISHA for low-energy and QGSJET for high-energy hadronic interactions. Primary particles are generated between 200 and $10^6$~GeV covering the angular domain of $0^0$ to $90^0$ for zenith angles and $0^0$ to $360^0$ for azimuthal angles with a power-law energy spectrum of index $-2.7$~\cite{pdg2024}. Each muon in a multiple-muon event is required to have an energy above \SI{54}{\giga\electronvolt}, the minimum needed to traverse the overburden above the NOvA ND. Muons incident at larger zenith angles require additional energy due to the longer rock depth. The surviving energy after propagation through rock of density $\rho$ and thickness $d$ is calculated using  
\begin{equation}\label{eq:energycorrection}
    E_\mu = (E_{\mu,0} + \epsilon) e^{-b d \rho} - \epsilon,
\end{equation}
where $E_{\mu,0}$ is the initial energy, $b = 4 \times 10^{-6}~\text{cm}^2/\text{g}$, and $\epsilon = 500$~GeV are standard-rock parameters~\cite{pdg}. Only muons with $E_\mu > 0$ are retained.

CORSIKA models atmospheric density with a five-layer parametrization. For the first four layers the mass overburden $X(h)$ decreases exponentially with height $h$ (from sea level),  
\begin{equation}\label{eq:xh1}
    X(h) = a_i + b_i e^{-h/c_i},
\end{equation}
while in the fifth layer, the dependence is linear,  
\begin{equation}\label{eq:xh2}
    X(h) = a_5 - b_5 h/c_5.
\end{equation}
Since there is no default profile for Fermilab, we extracted site-specific vertical atmospheric profiles from GDAS~\cite{GDAS} for 2017 and fitted them to this parametrization to obtain the coefficients $a_i$, $b_i$, and $c_i$.

Proton showers are generated with these atmospheric parameters and the simulation settings above. Cosmic muons from each shower are distributed uniformly across an extended $xz$ (horizontal surface where $z$ axis is along the neutrino beam direction) plane encompassing the detector ($x \in [-32,32]$~m, $z \in [-30,46]$~m). However, the NOvA ND spans $x \in [-2,2]$~m and $z \in [0,16]$~m; the extended surface ensures inclusion of $>99\%$ of muons intersecting the detector volume. For multiple-muon events, the first muon is placed randomly on this plane, while the others retain their relative spacings and orientations from CORSIKA. Showers producing at least one muon in the ND are propagated through the detector using GEANT4~\cite{geant4, geant42, geant43}. Upward-going neutrino-induced muons are not considered.

\begin{figure}
    \centering
    \includegraphics[width = 0.5\textwidth]{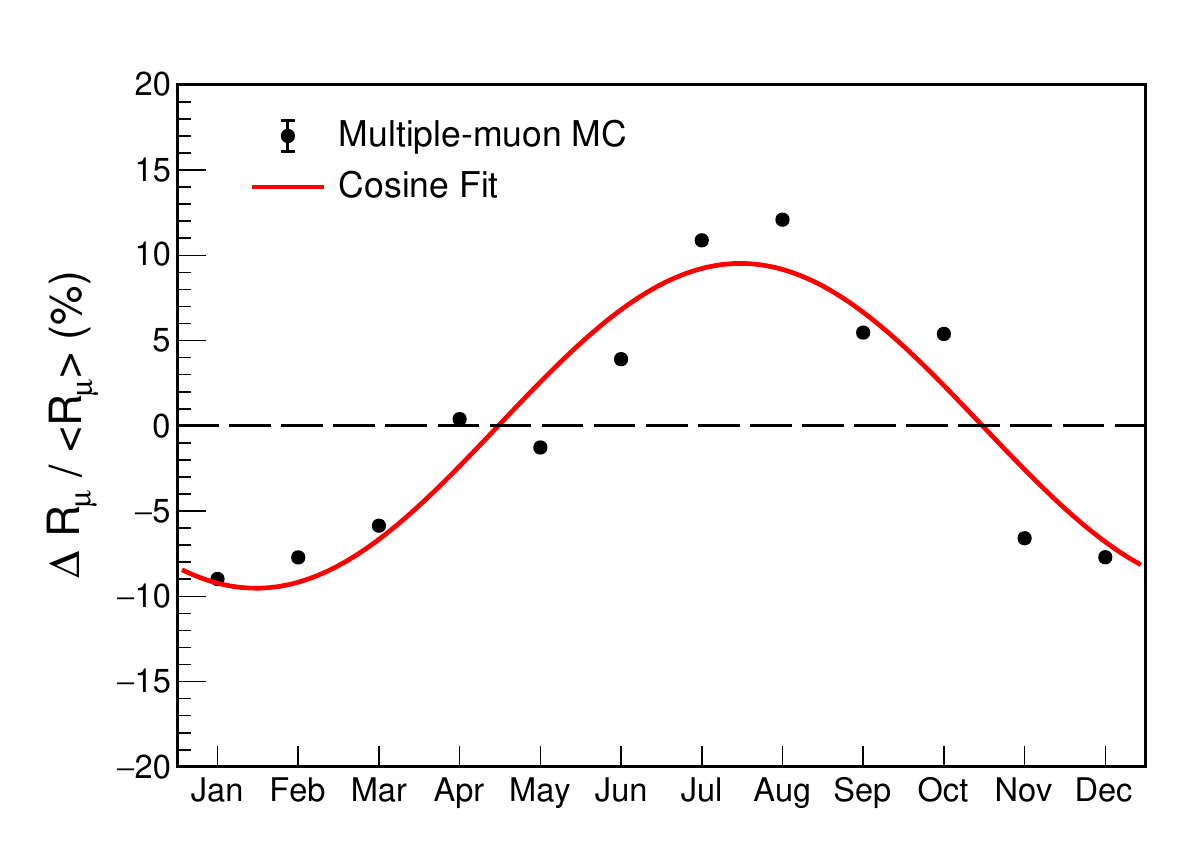} 
    \caption{Seasonal variation of multiple-muon rate for an infinite detector at the depth of the NOvA ND from CORSIKA initiated by primary proton cosmic rays. With detector size not included, the summer peak from atmospheric density effects on the muon rate is seen. The solid curve shows a cosine fit.}
    \label{fig:seasonalSurface}
\end{figure}

\begin{figure}[htp]
    \centering
    \includegraphics[width = 0.5\textwidth]{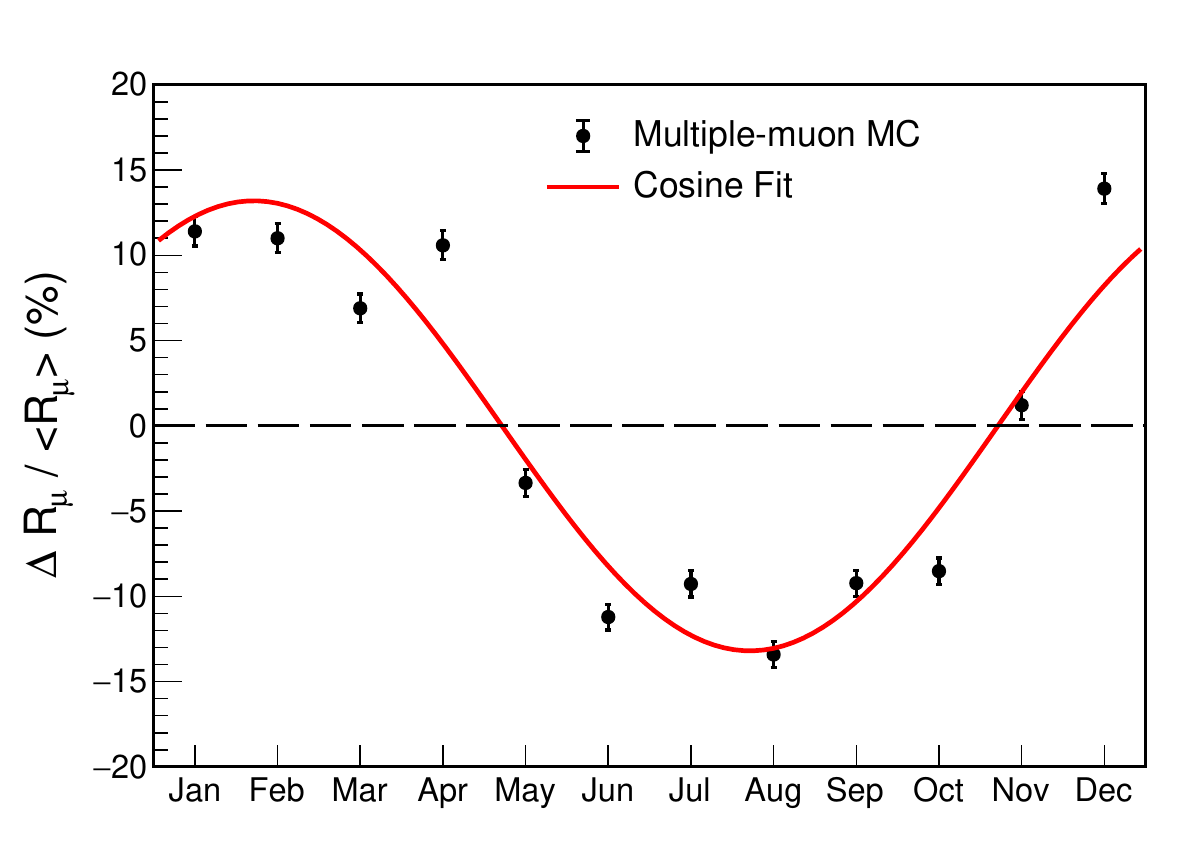}
    \caption{Seasonal variation of multiple-muon rate for an \SI{4}{\meter} by \SI{15.9}{\meter} detector at the depth of the NOvA ND from CORSIKA initiated by the primary proton cosmic rays. The solid curve shows a cosine fit. With a detector size less than the average of the muon shower at this depth (\SI{105}{\meter}), the summer peak has turned into a summer deficit because individual muons in the shower miss the detector, causing the event to not be classified as a multiple muon event. Error bars represent statistical uncertainties.}
    \label{fig:seasonalDet}
\end{figure}

Seasonal effects are studied with two complementary approaches. First, we consider all muons in a CORSIKA event with sufficient energy to reach the ND depth, corresponding to an ``infinite detector.'' Second, we simulate detector response with the standard NOvA ND software~\cite{ndxsec}. The simulated multiple-muon rate for an infinite detector shows a summer maximum as displayed in Fig.~\ref{fig:seasonalSurface}, opposite that seen in multiple-muon measurements from data. These simulated infinite-detector events were passed through the finite ND using GEANT4 simulation, and the same analysis cuts described in Ref.~\cite{novaND} and later in Sec.~\ref{sec: Cosmic muon data in the NOvA Near Detector} were applied. As a result, the time of the seasonal maximum shifted from summer to winter, as shown in Fig.~\ref{fig:seasonalDet}. This result from our simulation is consistent with previous observations (NOvA~\cite{novaND} and MINOS~\cite{minos}). The contrast between Fig.~\ref{fig:seasonalSurface} and ~\ref{fig:seasonalDet} strongly suggested that the difference was due to a geometrical effect related to the size of the detector, and not a result of differences in the meson content of the hadronic shower.

The CORSIKA implementation does not fully capture the absolute magnitude of single-muon seasonal variation, which is well understood theoretically and experimentally~\cite{SeasonalTh}. Nevertheless, the atmospheric density profile provides a reasonable basis for the relative seasonal differences in hadronic shower development. For the present analysis, only the comparison of the relative phase and modulation strength between single- and multiple-muon events in a fixed detector is used.

\section{Temperature data}
\label{sec: Temperature Data}
Atmospheric temperature data are taken from ERA5~\cite{era5}, the fifth-generation atmospheric reanalysis produced by the European Centre for Medium-Range Weather Forecasts (ECMWF)~\cite{ecmwf}. They provide temperature readings at 37 pressure levels, distributed nonuniformly from 1 to 1000 hPa. Temperatures are extracted from ECMWF within the grid area ranging from (41$^\circ$N, 87$^\circ$W) to (42.5$^\circ$N, 88.5$^\circ$W), which corresponds to the NOvA ND location.

\begin{figure}
    \centering
    \includegraphics[width = 0.5\textwidth]{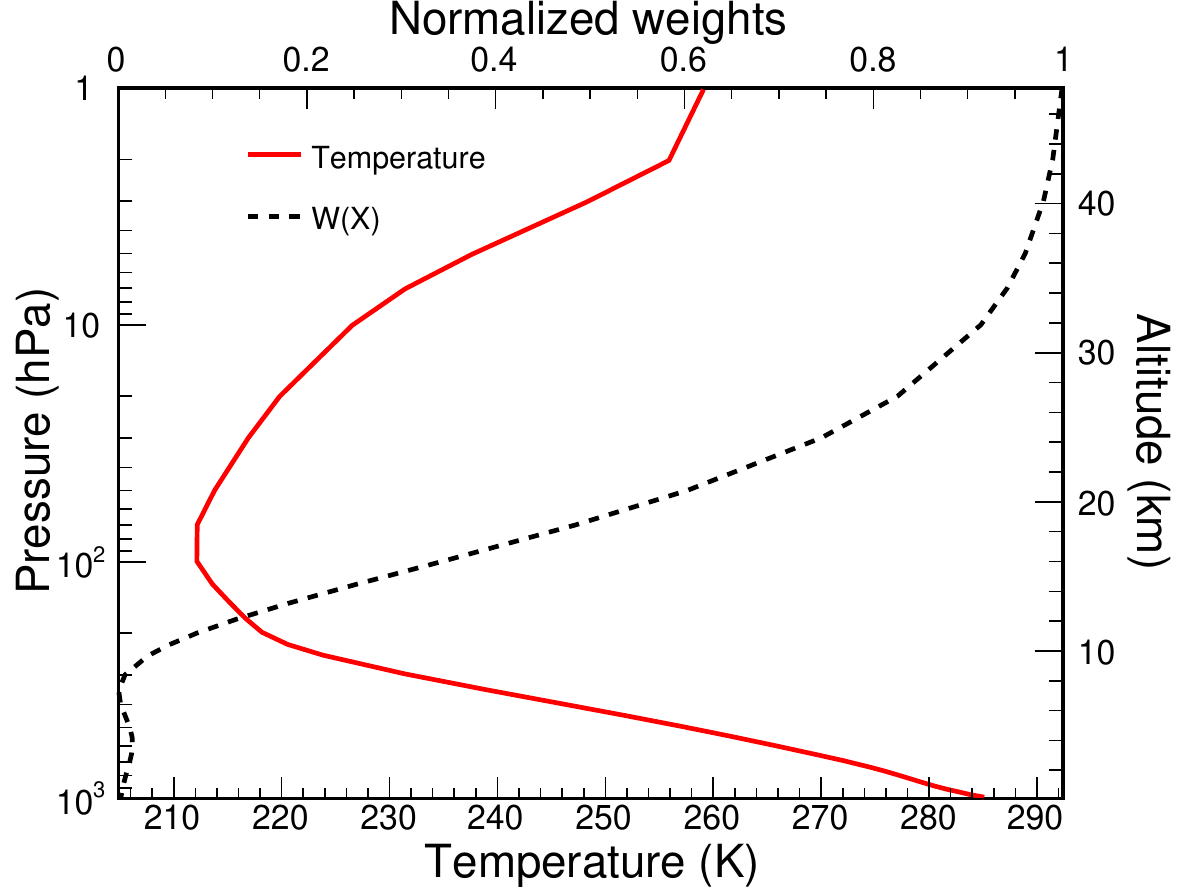}
    \caption{Variation of average temperature with altitude and total weight function used to calculate effective temperature. This weight corresponds to the fraction of cosmic ray primaries left at a given height: for example, about half have already interacted at a height of \SI{18}{\kilo\meter}.}
    \label{fig:tempweight}
\end{figure}

Since we do not know the exact altitude at which each muon is produced in the atmosphere, effective temperature is used instead of exact temperature. This is essentially a weighted average of the atmospheric temperature, where the weights reflect the probability distributions of meson (mainly pion and kaon) production and decay into muons at different altitudes. The effective temperature is defined as~\cite{SeasonalTh}
\begin{equation} \label{eq: Teff}
    T_{e\!f\!f} =\frac{\sum_{i=1}^{37} {T(P_i)[W_\pi(P_i) + W_K(P_i)]\Delta P_i}} {\sum_{i=1}^n[W_\pi(P_i) + W_K(P_i)]\Delta P_i}.
\end{equation}
 Here, $T(P_i)$ is the temperature at pressure level $i$ ($P_i$), and $W_K(P_i)$ and $W_\pi(P_i)$ are the weights responsible for the contribution as a function of pressure (altitude) for those $K$ and $\pi$ decays that make muons. The effective temperature is calculated using cross sections and lifetime values from~\cite{SeasonalTh}, along with temperature values obtained at 37 pressure levels from ECMWF. \autoref{fig:tempweight} shows the variation of ECMWF temperature data of the atmosphere over the Fermilab area, averaged over the four years 2018--2022, as a function of altitude and pressure. It also shows the total weight $W_\pi + W_K$ used in Eq.~\ref{eq: Teff}.

\section{COSMIC MUON DATA IN THE NO\lowercase{v}A NEAR DETECTOR}
\label{sec: Cosmic muon data in the NOvA Near Detector}
Cosmic muons in the NOvA ND are collected using a data-driven activity trigger. This trigger initiates data collection when there are at least ten hits in detector planes, requiring hits on at least three planes in each of the two views ($xz$ and $yz$) and a minimum of eight planes in total. Additionally, hits must be recorded in at least five planes out of six consecutive planes.

Collected hits are reconstructed using the Hough transform~\cite{houghtrans} method which is widely used for line reconstruction. All possible reconstructed tracks in 2D are stored. These 2D tracks from each plane are merged to create a 3D track.

To obtain a clean sample of cosmic-ray–induced multiple-muon events, additional selection criteria are imposed. First, a fiducial cut requires both the start and end points of each track to be within \SI{50}{\centi\meter} of the detector surface. This removes events in which electrons, produced nearly parallel to muons in the surrounding rock, generate short tracks in the detector, as well as contained or partially contained muon events. Second, each muon track must cross at least 10 planes to suppress short, poorly reconstructed tracks. Finally, to reduce contamination from single muon events occurring in close temporal proximity, the time difference between the first and any additional tracks in a multiple-muon event is required to be less than \SI{100}{ns}. The fiducial requirement rejects 16.9\% of events, the plane requirement removes 16.2\%, and the timing cut discards an additional 0.04\%. 

\begin{figure}
    \centering
    \includegraphics[width = 0.5\textwidth]{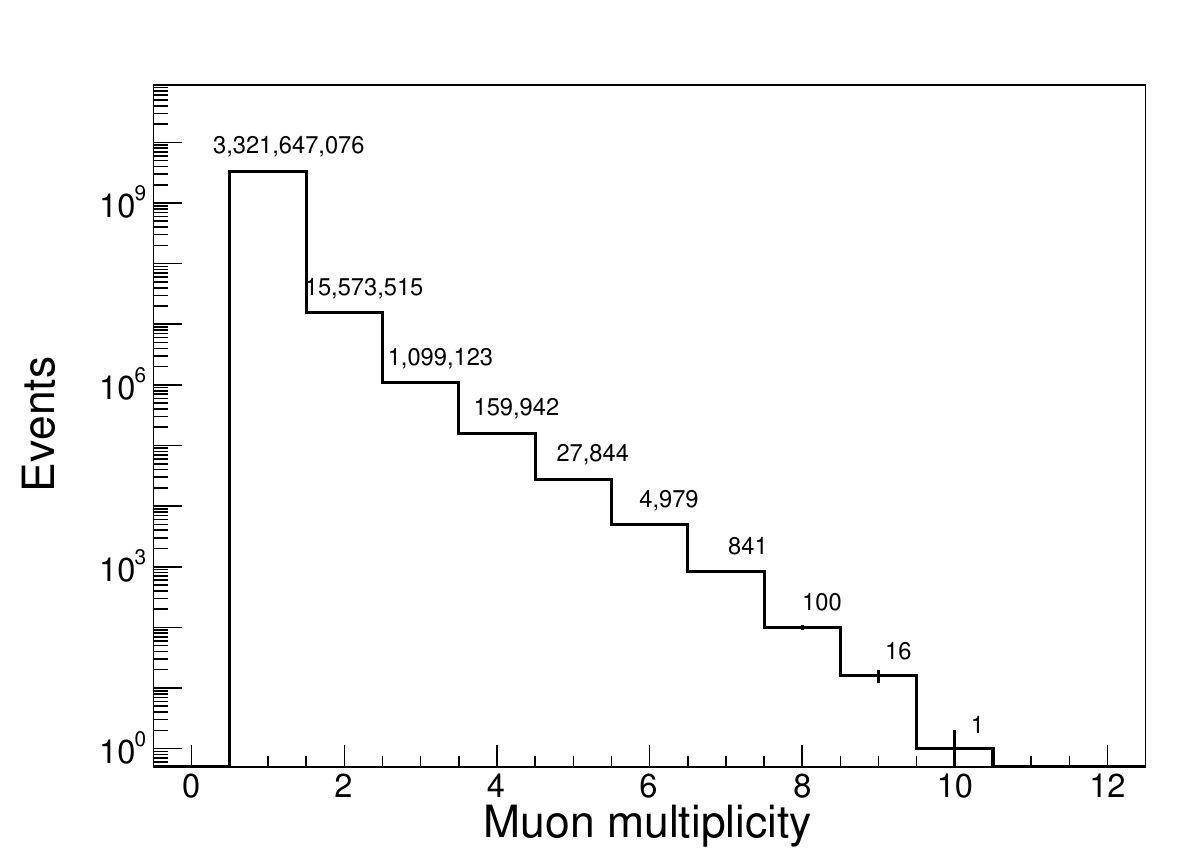}
    \caption{Distribution of number of reconstructed muon tracks in an event (multiplicity) in the dataset (2018--2022) used for this analysis. The highest multiplicity observed is 10. Error bars represent statistical uncertainties. }
    \label{fig:multiplicity}
\end{figure}

\begin{figure}
    \centering
    \includegraphics[width = 0.5\textwidth]{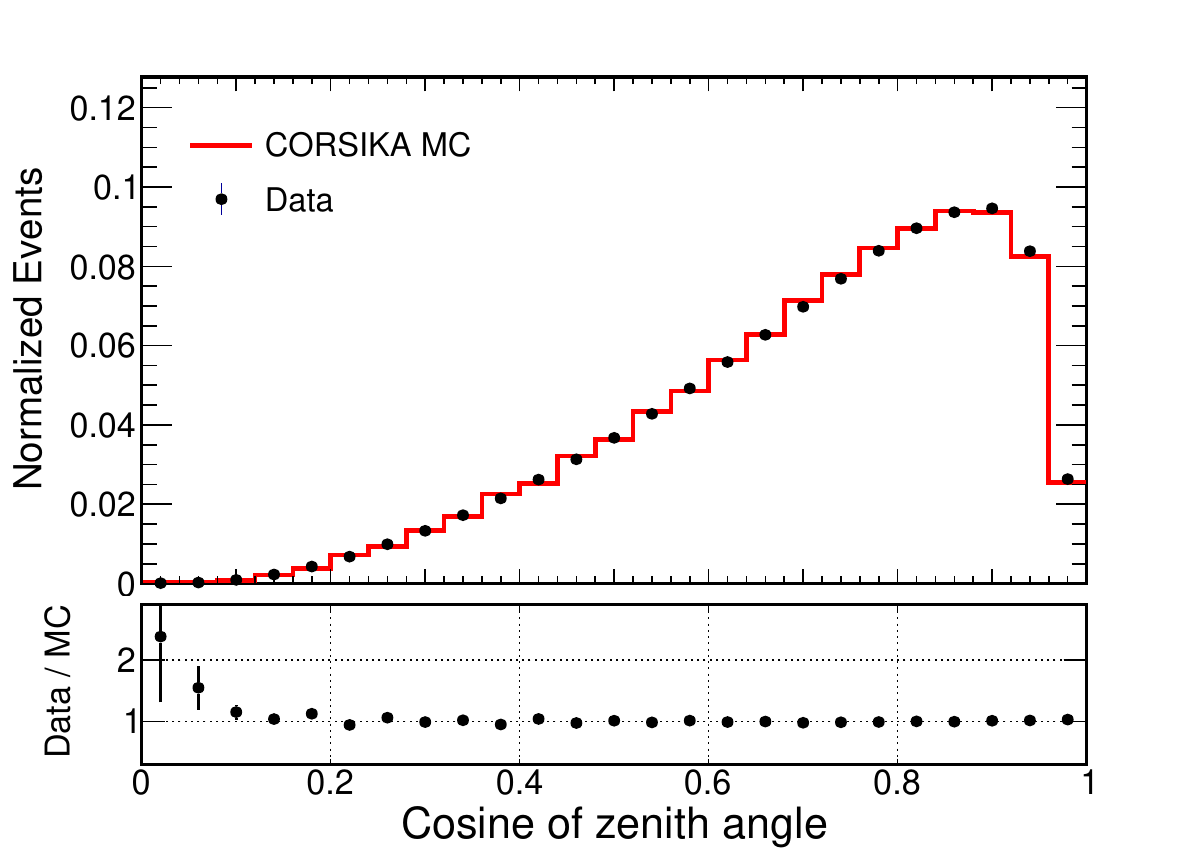}
    \caption{Distribution of the cosine of the angle between the track and the vertical upward direction (zenith angle). The ratio of data over the CORSIKA MC prediction is shown in the lower panel with statistical uncertainty.}
    \label{fig:datamc}
\end{figure}

The track multiplicity distribution of selected events from the ND data is shown in Fig.~\ref{fig:multiplicity}. The maximum multiplicity observed in the ND data is 10. The distribution of cosine of the zenith angle, representing the angle between the muon track and the vertical upward direction, is illustrated in Fig.~\ref{fig:datamc}. The distribution exhibits good agreement between data and Monte Carlo simulation.

\begin{figure}
    \centering
    \includegraphics[width = 0.48\textwidth]{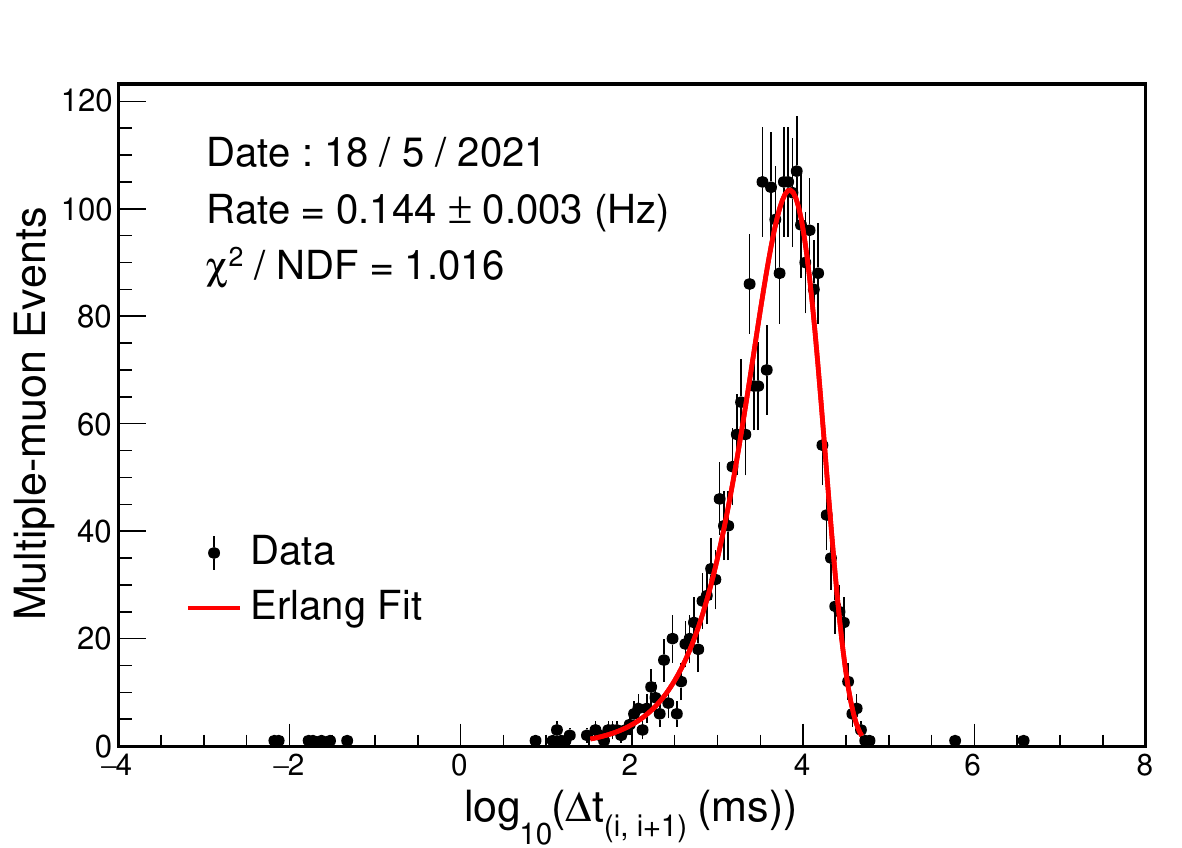}
    \caption{Logarithm (base 10) of time difference (in ms) between $i$ and $(i+1)$th multiple-muon events from one day of data. The distribution is fitted with an Erlang distribution in the range 1.5 to 4.7, yielding an observed multiple muon rate of $0.144 \pm 0.003$ Hz for that day. Selected noise events that do not match the hypothesis of a cosmic ray multiple muon of that rate are seen on the left with very short $\Delta t$. Long $\Delta t$'s coming from unaccounted for dead time are seen on the right of this plot. Error bars represent statistical uncertainties.}
    \label{fig:deltaT}
\end{figure}

\begin{figure}
    \centering
    \includegraphics[width = 0.48\textwidth]{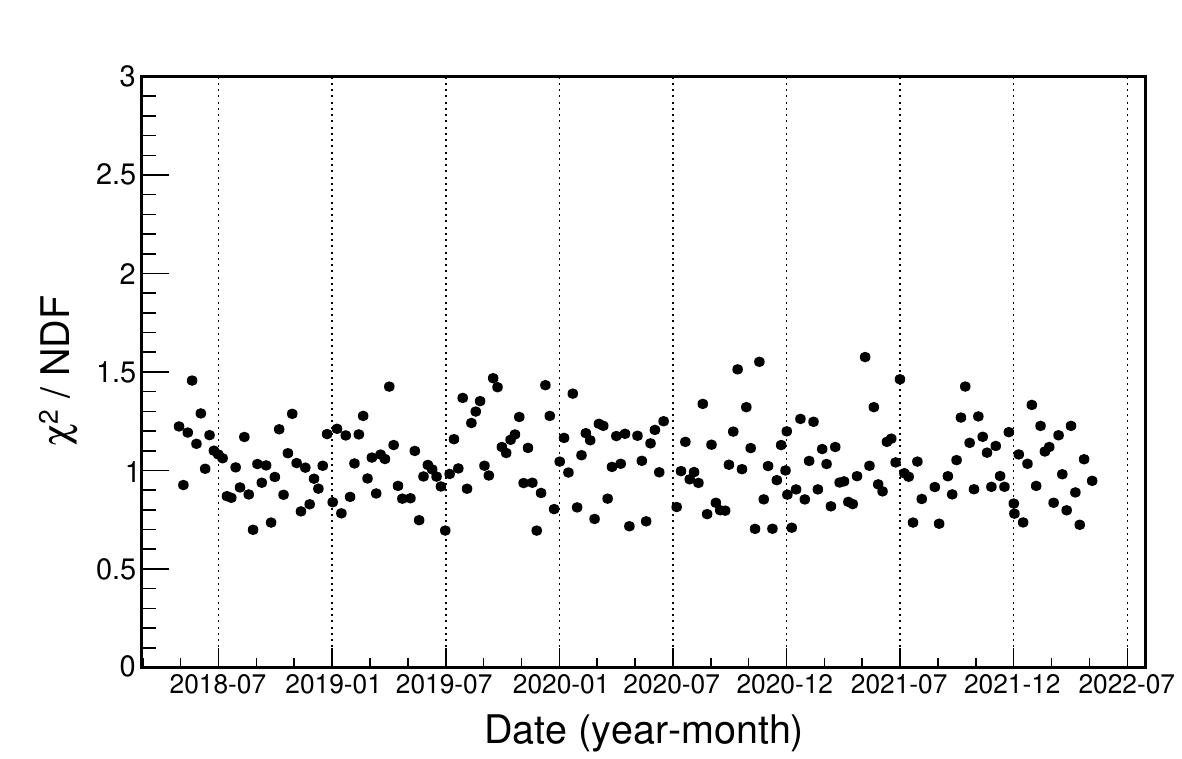}
    \caption{$\chi^2$/NDF distribution of Erlang fit for multiple-muon data in weekly binning. Since the dominant source of selected events are cosmic ray multiple muon showers following Poissonian arrival times, the Erlang fit used to extract the rate is well matched to the data and robust over time.}
    \label{fig:chisqsix}
\end{figure}

There are two differences between the multiple muon data in this analysis and that in ~\cite{novaND}. First, four years of new data are presented here. Second, the first analysis obtained a rate by simply dividing the number of selected events in a time period by the elapsed time covered by a data file. Both terms have uncertainties that, although small, can be difficult to quantify. For example, events that are not cosmic ray induced multiple muons might increase the event count, or unaccounted for dead time in the data acquisition might decrease the live time. Furthermore, the NOvA data is not stored in time order on disk, resulting in events that ``belonged" in one file being stored in another, changing counts in both files. This analysis uses a fit based on sorted arrival time differences to eliminate these issues. Hence, we first reordered all of the multiple-muons based on their arrival time. Then rates are obtained by performing a fit to the distribution of time difference (\(\Delta t\)) between consecutive multiple-muon events with an Erlang's probability density function (PDF)~\cite{forbes2011}. The analytical form of the Erlang PDF,
\begin{equation}
    f(\Delta t; k, \lambda) = \frac{\lambda^k {\Delta t}^{k-1} e^{-\lambda \Delta t}}{ (k-1)!}
\end{equation}
where $k$ is called the shape parameter and $\lambda$ is the rate parameter.
The shape parameter $k$ is equal to 1 for the two consecutive multiple-muon events. So, putting $k=1$ for consecutive events, the PDF becomes
\begin{equation}
    \label{eq:erlang2}
    f_t(\Delta t) = \lambda e^{-\lambda \Delta t}, \qquad \Delta t \geq 0.
\end{equation}
Now under the transformation \(x = \log_{10}(\Delta t)\), the Eq.~\ref{eq:erlang2} takes the form
\begin{equation}
    f_x(x) = A \, \lambda \, 10^{x} \, e^{-\lambda 10^{x}},
\end{equation}
where \(A\) is a normalization factor. This expression is the functional form used to extract the multiple-muon event rate from the \(\log_{10}(\Delta t)\) spectrum. Since $\Delta t$ is computed in $ms$, $\lambda$ is obtained in $ms^{-1}$, which is then converted into $Hz$.  A representative fit for a random week is shown in Fig.~\ref{fig:deltaT}, with the quality of fits across all weekly bins summarized in Fig.~\ref{fig:chisqsix}.

The principal advantage of this method is its robustness to irregularities in the data stream. Occasional detector dead times or DAQ interruptions can produce unusually large \(\Delta t\) values. In the \(\log_{10}(\Delta t)\) histogram, such events populate the far tail of the distribution and lie well outside the region around the peak that is used in the fit. Similarly, very small \(\Delta t\) values can arise from split tracks or double triggers in the DAQ. These appear well below the main peak and are again excluded from the fitting region. As a result, the fitted event rate is determined only from the central part of the distribution, where the data follow the expected exponential law that comes from the Poissonian process of the signal (multiple muons).  This allows calculation of error bars on the rates from pure statistics, by eliminating systematic errors in both the numerator and denominator of the naïve ``divide count by the live time of a file" process used in~\cite{novaND}.  



\begin{figure}
    \centering
    \includegraphics[width = 0.48\textwidth]{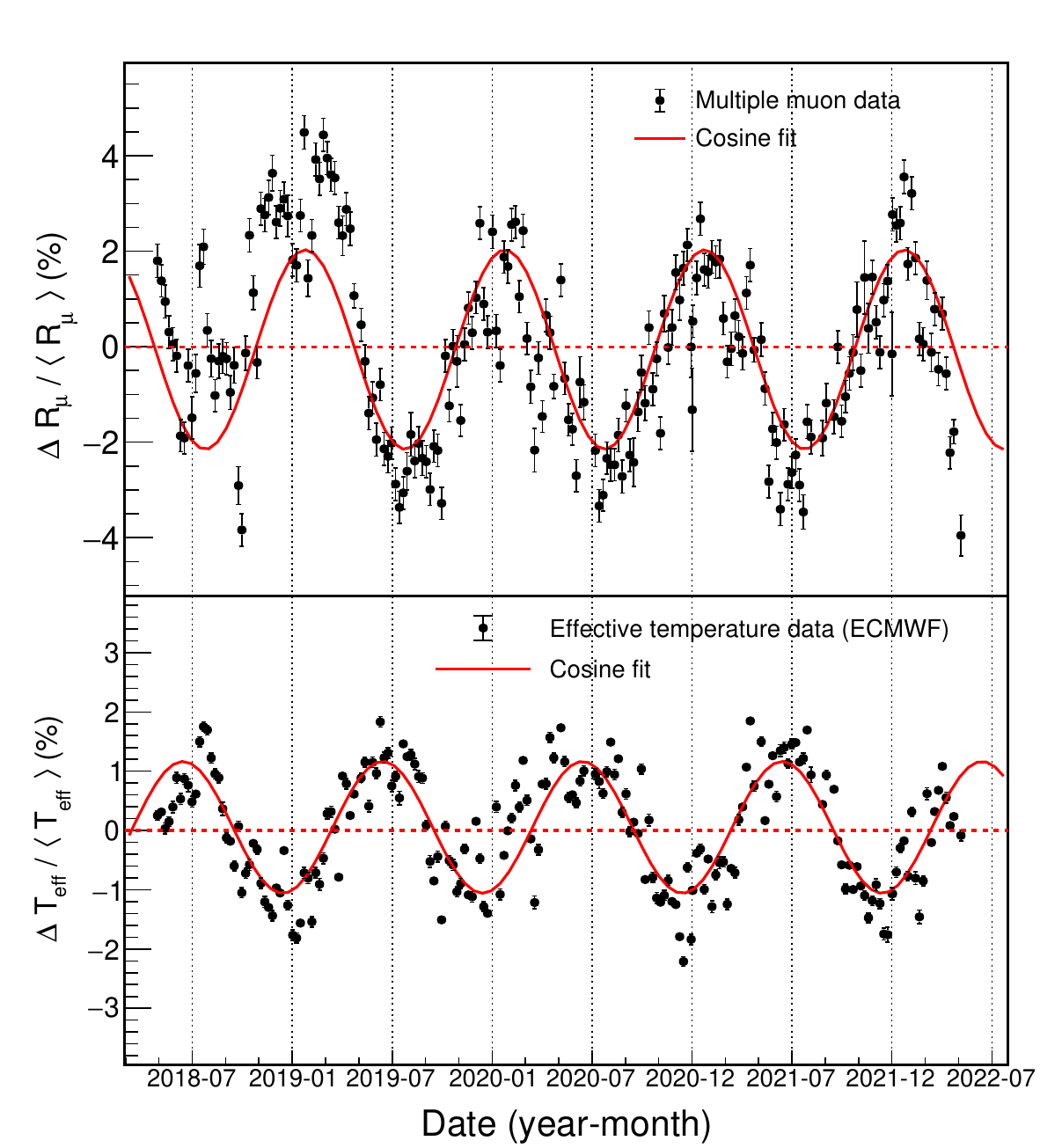} 
    \caption{Seasonal variation in weekly binning: The percentage variation of observed multiple-muon rate (top) with cosine best fit (solid curve) and the percentage variation of effective temperature (bottom) with cosine best fit (solid curve).}
    \label{fig:variationSix}
\end{figure}

Figure~\ref{fig:variationSix} shows the variation of the multiple-muon event rate seen in those four years of ND data, as well as the effective temperature. Both the data and the temperature are presented in weekly bins. The percentage variation is calculated using 
\begin{equation}
    \frac{\Delta R_\mu}{\langle R_\mu \rangle}  =  \frac{(R_\mu - \langle R_\mu \rangle)}{\langle R_\mu \rangle}
\end{equation}
where $R_\mu$ is the multiple-muon rate and $\langle R_\mu \rangle $ is the average rate over the four years of data. Systematic errors on the calculation of the rate, temperature and fit parameters are presented in~\cite{novaND}. None of these systematic errors varies as a function of season.

The muon rate in the multiple-muon data displays a seasonal trend that is out of phase with the effective atmospheric temperature. Although periodic, there is no \textit{a priori} reason to expect a cosine shape. However, it does serve to extract the phase of the magnitude maxima, so the percentage variations of multiple-muon rate and effective temperature are fitted with cosine curves, 
\begin{equation}
    V_0 + V \cos\left[{\frac{2\pi}{t_{year}}(t-\phi)}\right], 
\end{equation}
where $V_0$ is the average rate, $V$ is the magnitude, $t_{year}$ is the time period (fixed at one year) and $\phi$ is the phase. The fitted parameters are shown in \autoref{Tab: fitSinSixHour}. The multiple-muon rate varies by $\pm2\%$ in rate, and has the opposite phase as the temperature. The correlation coefficient between the rate of multiple muons and the effective temperature is found to be $-0.98 \pm 0.03$.

\begin{table}[h]
    \centering
    \caption{Best fit parameters for the cosine fit of percentage variation of the multiple-muon data and effective temperature}
    \begin{tabular}{c c c}
    \hline
     & Muon data & Temperature data  \\
    \hline
    $V_0$ (\%) & -0.060 $\pm$ 0.002 &  0.050 $\pm$ 0.006\\
    Magnitude ($V$) (\%) & 2.09 $\pm$ 0.04 & 1.11 $\pm$ 0.01\\
     Phase ($\phi$)  & 0.34$\pi$ $\pm$ 0.05$\pi$ & 1.67$\pi$ $\pm$ 0.12$\pi$\\
    \hline
    \end{tabular}
    \label{Tab: fitSinSixHour}
\end{table}

\section{Explanation}
\label{sec:explanation}
\subsection{The altitude-geometry explanation}
\label{sec:age}

\begin{figure}
    \centering
    \includegraphics[width = 0.30\textwidth]{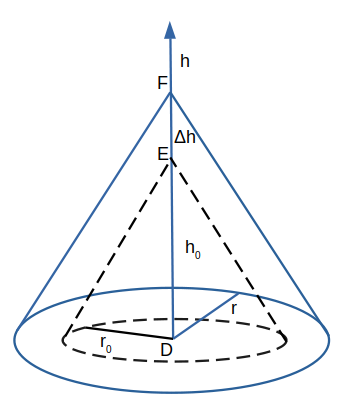}
    \caption{This shows how the radial spread of muons increases with increasing temperature. In summer, the atmosphere expands, so the muon birth altitude ($h$ = DF) increases compared to that in winter ($h_0$ = DE), and hence the radial spread also increases in summer ($r$) compared to winter ($r_0$) (figure from Ref.~\cite{decor_new}). }
    \label{fig:radialSpreadMuon}
\end{figure}

The total mass of the atmosphere remains approximately constant throughout the year, but seasonal temperature differences cause vertical expansion of atmosphere in summer and contraction in winter. Under the assumption of an isothermal atmosphere, the ideal gas law $PV = nRT$ implies that a $\pm$2\% change in absolute temperature results in a corresponding $\pm$2\% shift in altitude for a given pressure level. As shown in Fig.~\ref{fig:radialSpreadMuon}~\cite{decor_new}, for a shower origin at fixed pressure, the increased altitude (point F) in summer leads to a broader lateral spread of muons at the detector. In contrast, during winter, the lower altitude (point E) results in a more concentrated footprint.

\begin{figure}
    \centering
    \includegraphics[width = 0.5\textwidth]{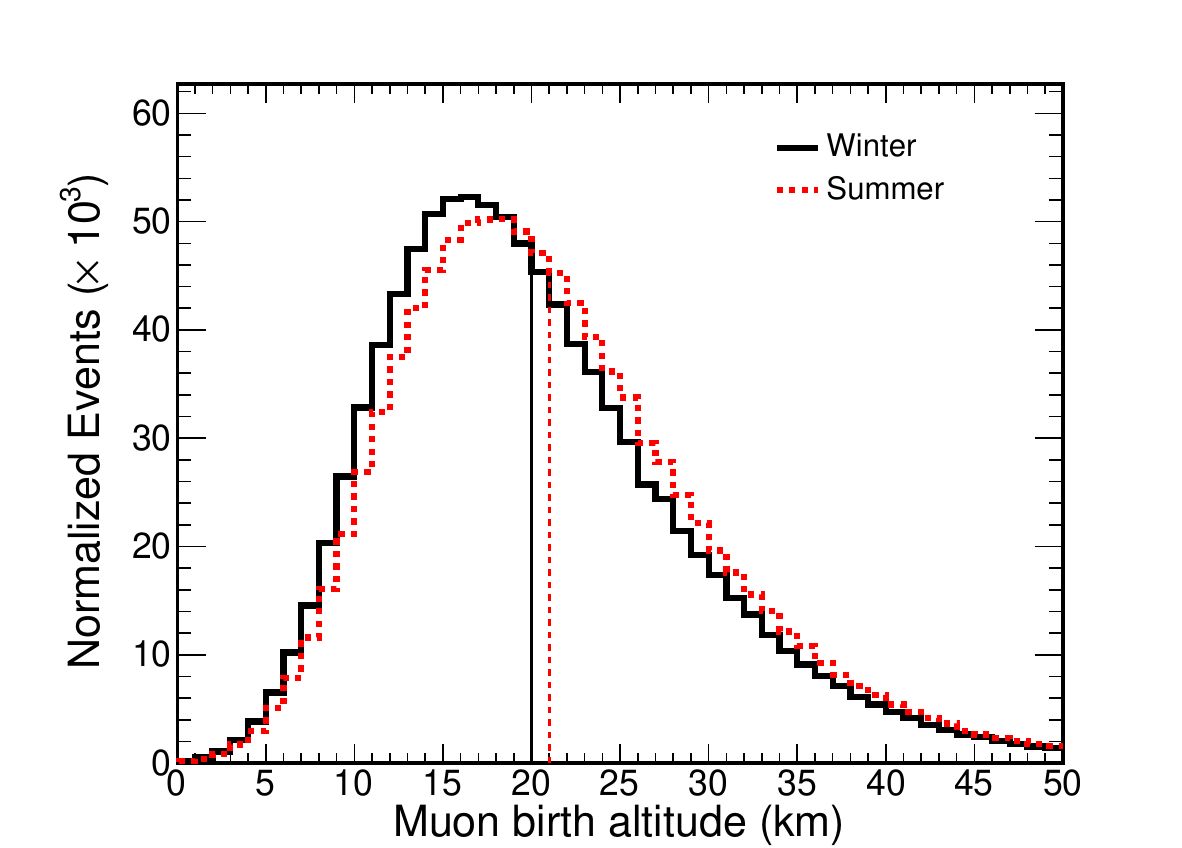} 
     \caption{Production altitude distribution of multiple-muon events from CORSIKA with muon threshold energy of \SI{54}{\giga\electronvolt}. Three winter months (December--February) and three summer months (June--August) are compared. Vertical lines represent the mean muon birth altitude for summer (red dashed line) and winter (black solid line).}
    \label{fig:muonalti}
\end{figure}

\begin{figure}
    \centering
    \includegraphics[width = 0.5\textwidth]{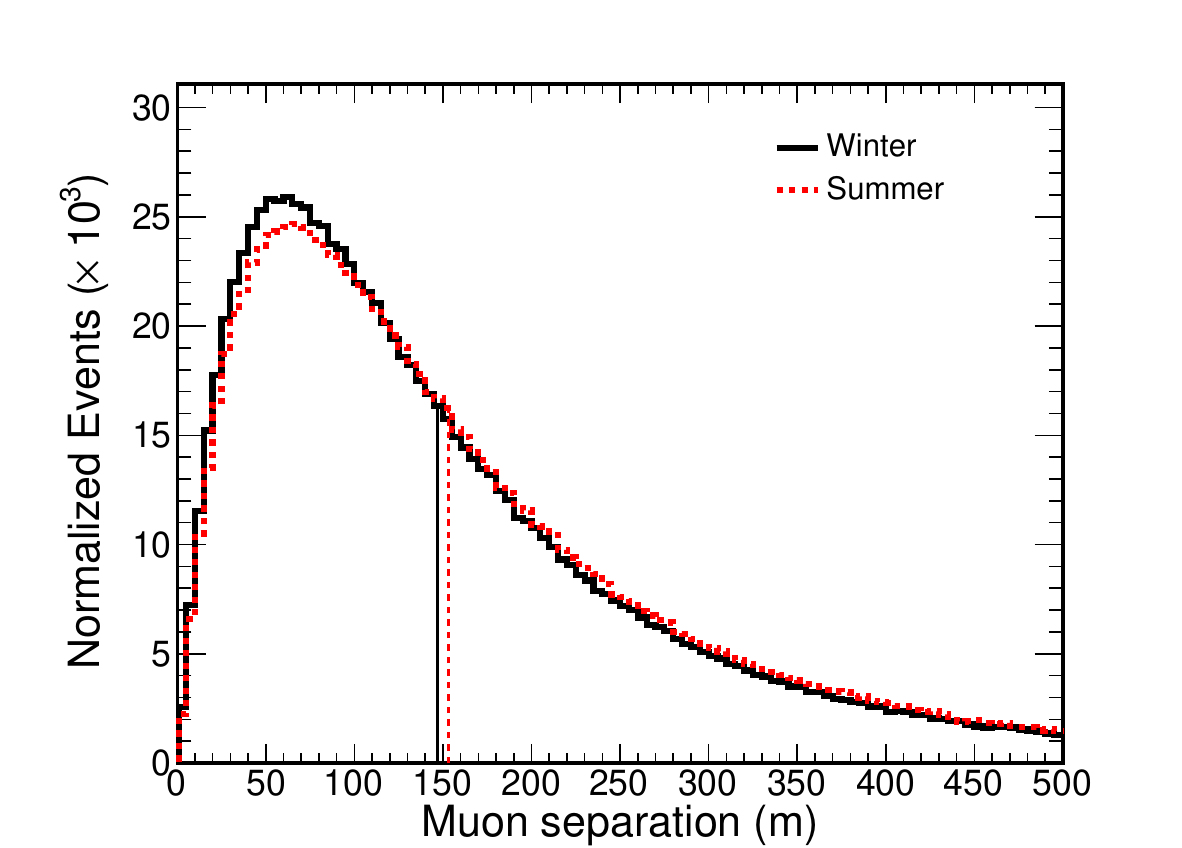} 
     \caption{Muon separation distribution of multiple-muon events from CORSIKA with muon threshold energy of \SI{54}{\giga\electronvolt}. Three winter months (December--February) and three summer months (June--August) are compared. Vertical lines represent the mean muon separation for summer (red dashed line) and winter (black solid line).}
    \label{fig:muonsep}
\end{figure}

We use the events generated with the CORSIKA simulation to study the altitude and muon separation variation during summer and winter.  Figure~\ref{fig:muonalti} shows the mean birth altitude of muons in the upper atmospehere from multiple muon events is higher in summer than in winter. As a result, the separation among muons for those events also increases in summer compared to winter, as shown in the Fig.~\ref{fig:muonsep}. It is instructive to consider the typical transverse momentum of mesons in hadronic showers, and how that impacts the separation distribution of muons reaching a detector. The opening angle between a secondary pion and the primary cosmic ray direction in an air shower is approximately proportional to $p_t / p_l$, where $p_t$ is the pion’s transverse momentum and $p_l \sim E_\pi \sim E^\mu$ is its longitudinal momentum. Consequently, the typical lateral separation between the two muons originating from different pions in the same shower can be expressed as
\begin{equation}
\label{eq:muonseparation}
\sqrt{2}\times h \times (p_t / E^\mu),
\end{equation}
where $h$ is the altitude at which the pions decay. Since only higher-energy muons can reach deeper underground detectors, the required $E^\mu$ increases with depth, leading to smaller lateral separations between muons for such underground detectors. To illustrate how the altitude-geometry effect depends on detector depth and muon energy, \autoref{tab:depthSeparation} shows the typical lateral separation of muons at various underground depths, calculated using Eq.~\ref{eq:muonseparation} for pions decaying at an altitude of 20 km and a typical pion $p_t$ of $300~\text{MeV/c}$. The corresponding minimum muon energies to reach the detector are calculated assuming standard rock with a density of $2.8~\text{g/cm}^3$ and a minimum ionizing energy loss of $2~\text{MeV}/(\text{g}\,\text{cm}^{-2})$. The muon separation decreases with increasing the energy (required to reach deeper detectors) and with increasing depth due to energy-dependent selection of muons closer to the shower core. 

\begin{table}[h]
\centering
\caption{Typical lateral separation of muons as a function of underground detector depth and muon energy threshold, calculated using Eq.~\ref{eq:muonseparation}. The separation indicates the typical transverse spread of muons from a cosmic ray shower at the detector level. }
\label{tab:depthSeparation}
\begin{tabular}{c c c}
\hline
Depth & $E^\mu_{min}$ & Muon  \\
(m)& (GeV) & Separation (m) \\
\hline
18    & 10    & 850 \\
80    & 50    & 170 \\
180   & 100   & 85 \\
800   & 500   & 17 \\
1800  & 1000  & 8.5 \\
\hline
\end{tabular}
\end{table}

For the NOvA or MINOS underground near detectors, with an energy threshold close to \SI{50}{GeV}, the typical muon separation is \SI{170}{\meter} which is much larger than the detector size. We can use \autoref{tab:depthSeparation} to determine whether any underground detector with a particular depth and area ($x\times z$) exhibits a winter or summer maximum. If the separation is much smaller than $x$ or $z$, a summer maximum will be seen. If the separation is larger or comparable, a winter maximum will be observed. However, the single muon rate remains unaffected by the altitude-geometry effect. For every single muon that misses a finite-sized detector because it starts at a higher altitude in the summer, another muon that would miss the detector in the summer now hits it. The muon flux does depend on the detector geometry.

The role of summer/winter differences in the origin altitude of muons in an air shower, and its effect on shower containment in a finite detector, was also considered before by MINOS~\cite{minos}.  However, the conclusion in the paper was based on the distributions of observed muon separation in the detectors, instead of the true separation between muons. That reference based its conclusion on the measured muon separation distribution, assuming that it was the true distribution. 

The dependence of the altitude–geometry effect as function of detector depth can also explain the seasonal behaviors observed in other experiments. For deeper detectors that require higher muon energy (TeV) thresholds, the altitude--geometry effect is minimal. The surface area of the MACRO experiment~\cite{macrodetector} is \SI{76}{m}$\times$\SI{12}{m} and the threshold energy required for an atmospheric muon to reach the detector is \SI{1.3}{TeV}. The average separation among muons in this case, is smaller than the size of the detector. Hence, MACRO observed a summer maximum for multiple-muon events~\cite{macromulti}.  In the MINOS FD \cite{minos}, the typical muon separation from an air shower is comparable to the size of the detector. So, the observed muon separation determines whether there is a summer maximum or a winter maximum. Calculations by members of the IceCube collaboration based on the altitude-geometry effect seemed to explain the MINOS and NOvA multiple-muon seasonal data~\cite{Gaisser:2021cqh, Gaisser:2021bwj}. The altitude-geometry effect was considered responsible for the seasonal results of the DECOR experiment \cite{decor}, which measured muons close to the surface of the Earth.  For those muons, with an energy typically of a few GeV, other effects such as pressure changes and muon decay are important, as discussed further in Sec.~\ref{sec:other}.
 


\subsection{Other seasonal explanation}
\label{sec:other}
The MINOS result in Ref.~\cite{minos} disfavored dimuon decays or temperature effects as contributing to the explanation of multiple-muon variation. However, it  considered the anticorrelation between the primary meson interaction and decay as the most plausible explanation for the observed phenomenon.
According to this hypothesis, higher temperatures in summer increase the decay probability of mesons, leaving fewer mesons available to inelastically scatter and produce additional secondaries in subsequent generations of the hadronic shower. To test this independently of CORSIKA, we used a simulation of hadronic showers in the atmosphere with densities corresponding to summer and winter~\cite{guan2023montecarlosimulationsfactors}.  We followed the meson and muon content in the hadronic shower as subsequent generations of both mesons and nucleons interacted, as a function of primary cosmic ray energy.  The simulation showed that a shower produces more muons when the primary meson decays rather than interacts, contrary to the hypothesis favored by MINOS~\cite{minos}.  In addition, this explanation would predict a winter maximum in an infinite detector, in contradiction with the CORSIKA result in Fig.~\ref{fig:seasonalSurface}.

\begin{table}[h]
    \centering
    \caption{Analytic calculation of the percentage of muons that reach sea level before they decay as a function of muon energy. $E_\mu$ is the muon energy, $N_0$ is the initial number of muons, and $N$ is the number of muons that survive. The summer/winter altitude difference is assumed to be 2\%.}
    \begin{tabular}{c c c c}
     \hline
     $E_\mu$ (GeV) &  $N/N_0$ (\%) & $N/N_0$ (\%) & Difference (\%)\\
      & (winter) & (summer) & \\
     \hline
      5 & 62.05 & 61.46 & 0.59 \\
      
      50 & 95.34 & 95.25 & 0.09 \\
     
      500 & 99.52 & 99.51 & 0.01 \\
      \hline
    \end{tabular}  
    \label{Tab: muondecay}
\end{table}

Another potential explanation for the observed seasonal pattern of multiple muons is muon decay, which was not discussed in Ref.~\cite{minos}. Due to relativistic effects, muon decay is more significant at lower muon energies, and thus its impact is expected to be largest for detectors located at or near the surface. An analytical estimate of the seasonal modulation due to muon decay for various energy thresholds is summarized in Table~\ref{Tab: muondecay}. Surface detectors such as the NOvA FD~\cite{novafd}, DECOR~\cite{decor, Tolkacheva2011}, and GRAPES~\cite{grapes3}, all with muon thresholds near 5 GeV, can exhibit a 0.6\% excess of events during the winter. In contrast, detectors such as the NOvA ND and MINOS ND, which have muon thresholds close to 50 GeV, would only show a smaller 0.09\% excess in winter, which is much smaller than the observed winter excess, and also inconsistent with Fig.~\ref{fig:seasonalSurface}. Muon decay is even less important for the MINOS FD and MACRO.

\section{Conclusion}
\label{sec:conclusion}
The NOvA ND multiple-muon event rates have been measured using the Erlang fitting technique, which eliminates systematic effects due to detector and DAQ instabilities. The measured multiple muon rate for the 2018 to 2022 data period is found to be anticorrelated with the effective temperature of the atmosphere. This anticorrelation is consistent with previous NOvA ND and MINOS measurements.

The altitude-geometry effect is the most relevant explanation for the opposite seasonal behaviors between single and multiple-muon events. Evidence supporting this conclusion comes from a comparison between the typical muon separation distribution as a function of underground detector depth \textit{vis-à-vis} the detector size, using the CORSIKA simulation, which demonstrates a summer maximum for multiple-muon events in an infinite detector. However, for the finite-sized NOvA or MINOS ND, the simulation shows a winter maximum for multiple-muon events. The effect of muon decay for \SI{54}{\giga\electronvolt} muons was found to be too small. In general, the altitude-geometry effect leads to a winter maximum for multiple muons in underground detectors if the muon separation is similar to or larger than the size of the detector, but when it is much smaller than the detector size, a summer maximum is to be expected. This agrees with observations from the MINOS FD, MINOS ND, NOvA ND, and MACRO. In particular, the results of Ref.~\cite{minos} and~\cite{novaND} are now understood.

\begin{acknowledgments}
This document was prepared by the NOvA collaboration using the resources of the Fermi National Accelerator Laboratory (Fermilab), a U.S. Department of Energy, Office of Science, HEP User Facility. Fermilab is managed by Fermi Forward Discovery Group, LLC, acting under Contract No. 89243024CSC000002.  This work was supported by the U.S. Department of Energy; the U.S. National Science Foundation; the Department of Science and Technology, India; the European Research Council; the MSMT CR, GA UK, Czech Republic; the RAS, the Ministry of Science and Higher Education, and RFBR, Russia; CNPq and FAPEG, Brazil; UKRI, STFC and the Royal Society, United Kingdom; and the State and University of Minnesota.  We are grateful for the contributions of the staffs of the University of Minnesota at the Ash River Laboratory, and of Fermilab. For the purpose of open access, the author has applied a Creative Commons Attribution (CC BY) license to any Author Accepted Manuscript version arising.
\end{acknowledgments}

\bibliography{multimuref} 
\bibliographystyle{apsrev4-1} 

\end{document}